\documentclass[a4paper,10pt]{article} 

\usepackage{opticameet3} 

\usepackage{fancyhdr}
\newcommand\authormark[1]{\textsuperscript{#1}}

\usepackage{amsmath,amssymb}
\usepackage[colorlinks=true,bookmarks=false,citecolor=blue,urlcolor=blue]{hyperref} 
\usepackage{dsfont,color,float,empheq}
\usepackage[table]{xcolor} %
\usepackage{graphicx,wrapfig,multicol,multirow}
\usepackage{psfrag,wrapfig}
\usepackage[font=footnotesize]{caption}
\usepackage{pgfplots}
\usepackage{subcaption}
\usepackage{cite}
\usepackage{setspace}
\usepackage{multicol}
\usetikzlibrary{patterns}
\usepackage{tikz,pgfplots}
\usetikzlibrary{positioning,shadows,shapes,backgrounds,calc}
\usepackage{pgfplotstable}

\begin{document}

\title{A Simple BER Expression for FSO Systems with Weak Turbulence and Pointing Errors}


\author{Carmen Álvarez Roa\authormark{*},  Yunus Can Gültekin, Kaiquan Wu, Cornelis Willem Korevaar and Alex Alvarado}

\address{
Department of Electrical Engineering, Eindhoven University of Technology, Eindhoven, The Netherlands
}

\email{\authormark{*}c.alvarez.roa@tue.nl} 

\begin{abstract}
\textit{We develop a simple approximation for the average BER for an FSO system impacted by weak turbulence and pointing errors. Numerical results show that the proposed expression accurately predicts the true BER.}
\end{abstract}


\vspace{-1ex}
\section{Introduction}
\vspace{-1ex}
Free-space optics (FSO) is currently considered a promising solution to establish high-capacity point-to-point wireless links thanks to its large available bandwidth \cite{5G}. 
FSO provides higher security, requires no spectrum licensing, and generally has a lower cost~\cite[Sec.~I-B]{Survey}. 
However, FSO links are heavily affected by atmospheric turbulence, pointing errors, and atmospheric losses \cite{Farid2007}. 
The channel model studied in this work includes these three effects as the main phenomena deteriorating the performance of an FSO communication system.

In the weak turbulence regime, the joint probability density function (PDF) that characterizes the channel gain coefficient determined by atmospheric turbulence, pointing errors, and atmospheric losses was developed in~\cite[Sec.~III-D]{Farid2007}. 
This PDF is based on the complementary error function. 
The integral form of this function makes it difficult to find simple expressions of important performance metrics such as the bit error rate (BER). 
In particular, the exact analytical average BER expression for on-off keying (OOK) signaling over an FSO channel involves a double integral that is complex to evaluate.  
Recently, we proposed in \cite{ICTON} two approximations for this BER expression that can be easily computed. 
However, these approximations have a certain inaccuracy that varies depending on the region of operation in terms of the strength of turbulence and pointing errors. 
For instance, in the high-pointing-error regime and for BERs between $10^{-2}$ and $10^{-5}$, the slope of the approximate BER curve in \cite[Fig.~2~(right)]{ICTON} is visibly different than that of the actual slope obtained via Monte Carlo simulations. 

This paper presents an analytical BER approximation for OOK signaling over an FSO channel that is much more accurate than our previous work~\cite{ICTON}. 
We show that our proposed BER approximation is in excellent agreement with Monte Carlo simulations.
This approximation also provides the correct BER slope in all regions of operation. 
Although our previous approximation in~\cite{ICTON} is computationally the simplest, the proposed expression here involves two simple integrals, and thus, it is still easier to compute than the exact expression.  
To the best of our knowledge, the expression we provide in this paper is the most accurate approximation of the average BER for an FSO channel affected by weak atmospheric turbulence, pointing errors, and atmospheric losses.

\vspace{-1ex}
\section{The Channel Model and the Proposed Approximation}
\vspace{-1ex}
We consider a traditional FSO communication system with intensity modulation and direct detection (IM/DD) using OOK modulation. The transmitter first modulates equiprobable OOK data symbols $x \in\{0, 2P\}$ onto the instantaneous intensity of an optical beam, where $P$ represents the average transmit optical power. The signal is then transmitted through the FSO channel and detected by a photodetector at the receiver side. The received photocurrent is directly proportional to the incident optical power through the detector responsivity $\eta$.

The received signal exhibits a randomly fluctuating intensity due to additive white Gaussian noise with variance $\sigma_n^2$,
as well as the random attenuation of the propagation channel modeled by the multiplicative channel gain $h$. In our model, this attenuation is due to three factors: atmospheric turbulence ($h_a$), pointing errors ($h_p$), and atmospheric loss ($h_l$). Since these phenomena create independent multiplicative losses, the channel gain coefficient is given by~\cite[Eq.~(2)]{Farid2007}
\vspace{-1ex}
\begin{equation}
    h=h_ah_ph_l.
    \label{Coeff_H}
    \vspace{-1ex}
\end{equation}
Note that $h_l$ is given by $\exp(-\sigma L)$~\cite[Sec.~III]{Farid2007}, where $\sigma$ (dB/km) is the attenuation coefficient and $L$ (km) is the link distance.
Moreover, $h_p$ and $h_a$ are random variables (RVs) whose distributions are discussed in \cite[Sec.~2]{ICTON}.

In the weak turbulence regime, the PDF that characterizes the complete channel gain $h$ in~\eqref{Coeff_H} is~\cite[Sec.~III]{Farid2007} 
\vspace{-1ex}
\begin{equation}
     f_{H}(h)=\frac{\gamma^2}{2(A_0 h_l)^{\gamma^2}} h^{\gamma^2 -1} \hbox{erfc} \left( \frac{\hbox{ln}\left( \frac{h}{A_0h_l} \right) + \mu}{\sqrt{8 \sigma_X^2}}\right) \exp{\left[2\sigma_X^2 \gamma^2 (1+\gamma^2)\right]} \;, \quad  \hspace{1ex} \mathrm{erfc}(z)\triangleq\frac{2}{\sqrt{\pi}}\int_{z}^{\infty} \exp{(-t^2)} dt, 
     \label{eq:fh_LN_sim}
   \end{equation}
where $\mu \triangleq 2\sigma_X^2 (1 + 2\gamma^2)$. In \eqref{eq:fh_LN_sim}, $\sigma_X^2$ is related to the weak turbulence model~\cite[Sec.~III-A]{Farid2007} and is proportional to the Rytov variance $\sigma_R^2$. $A_0$ is the fraction of collected power when there are no pointing errors, and is a function of the radius $a$ of the receiver aperture and the beam waist $\omega_z$, while $\gamma$ is the ratio between the equivalent beam radius and the pointing error displacement standard deviation $\sigma_s$ at the receiver. This last parameter is estimated as $\sigma_s\approx\theta_sZ$, where $\theta_s$ is the jitter angle.

The PDF \eqref{eq:fh_LN_sim} is a non-closed-form expression since it involves an integral, often evaluated numerically.
In this work, we focus on the BER after hard-decision (HD) detection based on the maximum likelihood rule. 
The \emph{exact} average BER $\bar{P_\text{b}}$~\cite[Eq.~(15)]{ICTON} can be obtained by integrating the BER conditioned on the channel gain $h$, $P_\text{b}(h) =\frac{1}{2}\hbox{erfc}\bigl( {\eta P h}/{\sqrt{2\sigma_n^2}}\bigr)$,
over the PDF in~\eqref{eq:fh_LN_sim}. It was shown in \cite[Eq.~(16)]{ICTON} that the exact average BER  is
\begin{equation}
      \bar{P_\text{b}}=\int_{0}^{\infty}  P_\text{b}(h)f_H(h)\, dh=\frac{\gamma^2}{\pi(A_0 h_l)^{\gamma^2}}\exp{[2\sigma_X^2 \gamma^2 (1+\gamma^2)]} \int_{0}^{\infty}  h^{\gamma^2 -1} \left[\int_{u}^{\infty} \exp{(-t^2)} \, dt  \int_{v}^{\infty} \exp{(-t^2)}\, dt\right]\, dh,
        \label{eq:BER_LN}
\end{equation}
where $u={\eta P h}/{\sqrt{2\sigma_n^2}}$ and $v={\left(\hbox{ln}\left( \frac{h}{A_0h_l}\right) + \mu\right)}/{\sqrt{8\sigma_X^2}}$. 
From~\eqref{eq:BER_LN}, we can see that the exact expression involves the integral of the product of two integrals, which is quite complex to evaluate.
In this work, we propose the following analytical BER approximation for OOK where $\hat{h}\triangleq A_0h_l \exp{(-\mu)}$:
\vspace{-1ex}
 \begin{align}  
      \bar{P_\text{b}}\approx\frac{\gamma^2}{2\sqrt{\pi}(A_0 h_l)^{\gamma^2}}\exp{[2\sigma_X^2 \gamma^2 (1+\gamma^2)]}  \Bigg[\int_{0}^{\hat{h}} 
      h^{\gamma^2 -1} \Biggl[1+& \dfrac{\exp{\left(\frac{-2\pi v}{\sqrt{6}}\right)}-1}{\exp{\left(\frac{-2\pi v}{\sqrt{6}}\right)}+1} \Biggr] 
      \frac{\exp{\left(-u^2\right)}}{u+\sqrt{u^2+\frac{4}{\pi}}}\, dh \nonumber\\
      & + \frac{2}{\sqrt{\pi}} \int_{\hat{h}}^{\infty} h^{\gamma^2 -1} \frac{\exp{\left(- v^2\right)} }{ v +\sqrt{ v^2 + \frac{4}{\pi}}} \frac{\exp{\left(-u^2\right)}}{u+\sqrt{u^2+\frac{4}{\pi}}} \, dh \Biggr].
    \label{eq:BER_Approx2_UpperBound}
    \vspace{-2ex}
 \end{align}
Since~\eqref{eq:BER_Approx2_UpperBound} only involves two one-dimensional integrals, this approximation is easier to compute than \eqref{eq:BER_LN}. 
To obtain \eqref{eq:BER_Approx2_UpperBound}, we applied the following approximation to~\hbox{erfc(z)}:
\vspace{-2ex}
 \begin{align}
 \hbox{erfc}(z)\approx & 
   \begin{cases}
    \dfrac{2}{\sqrt{\pi}}  \dfrac{\exp(-z^2)}{z+\sqrt{z^2+\frac{4}{\pi}}}, & \text{if}~z\geq0,\\ 
      1+\dfrac{\exp{\left(-\frac{2\pi z}{\sqrt{6}}\right)}-1}{\exp{\Big(-\frac{2\pi z}{\sqrt{6}}\Big)}+1}, &  \text{if}~z<0.
     \end{cases}
     \label{eq:Erfc_Aprox2}
     \vspace{-2ex}
 \end{align}
For the case $z \geq 0$ in \eqref{eq:Erfc_Aprox2}, we use the asymptotically tight (as $z\rightarrow \infty$) approximation in \cite[Eq.~(7.1.13)]{QFun}. On the other hand, when $z<0$, we use another asymptotically tight approximation (as $z\rightarrow -\infty$) given in \cite{QFun2}.

In~\cite{ICTON}, by applying a different approximation to \hbox{erfc(z)}, i.e., \cite[Eq.~(13)]{ICTON}, we derived the following even simpler average BER approximation:
\vspace{-2ex}
\begin{align}
\bar{P_\text{b}}\approx\frac{\gamma^2\sigma_X \sigma_n}{\eta P\pi(A_0 h_l)^{\gamma^2}} \exp{[2\sigma_X^2 \gamma^2 (1+\gamma^2)]}  \int_{\hat{h}}^{\infty} \frac{h^{\gamma^2 -2}}{\hbox{ln}(\frac{h}{A_0h_l})+\mu} \exp{\left(-u^2\right)} \exp{\left(-v^2\right)}  \, dh.
\label{eq:BER_Aprox2_ICTON}
    \vspace{-2ex}
 \end{align}
 Since~\eqref{eq:BER_Aprox2_ICTON} only considers an approximation of \hbox{erfc(z)} that applies only for positive arguments (\mbox{$z \geq 0$}), it only involves a single one-dimensional integral, which makes \eqref{eq:BER_Aprox2_ICTON} simpler to evaluate than~\eqref{eq:BER_Approx2_UpperBound}. However, as we will see in the next section,~\eqref{eq:BER_Aprox2_ICTON} presents some inaccuracies, while~\eqref{eq:BER_Approx2_UpperBound} is extremely accurate. 
Table \ref{tab:BER_Comparation} shows a comparison between the exact average BER~\eqref{eq:BER_LN}, and the approximations~\eqref{eq:BER_Approx2_UpperBound} and~\eqref{eq:BER_Aprox2_ICTON} in terms of accuracy and simplicity. 
In the next section, we compare the accuracy of these BER expressions for realistic link conditions. 
 \vspace{-1ex}
 \begin{table}[h]
        \centering
        \small
        \caption{Comparison of Average BER Expressions}
        \vspace{-7pt}
        \scalebox{1}{
        \begin{tabular}{cccc}
            \hline
            \cellcolor{black!15}{\textbf{Expression}} & \cellcolor{black!15}{\textbf{Ref.}} & \cellcolor{black!15}\textbf{{Accuracy}} & \cellcolor{black!15}{\textbf{Simplicity}}\\
            \hline
            \eqref{eq:BER_LN}    & \cite{ICTON} & Exact & Complex\\
            \eqref{eq:BER_Approx2_UpperBound}& \textbf{This work}& More accurate & Moderately simple \\
            \eqref{eq:BER_Aprox2_ICTON}  & \cite{ICTON}  &  Less accurate & Very simple\\
            \hline
        \end{tabular}
        }
        \label{tab:BER_Comparation}
\end{table}
\vspace{-2ex}
\section{Numerical Results}
\vspace{-1ex}
In Fig.~\ref{Fig:Main}, average BERs are plotted as a function of the average transmit power $P$. Table \ref{tab:sim_param} summarizes the parameters considered in our numerical analysis. These parameters are obtained from \cite{Farid2007,Korevaar03,Boluda2017}. In terms of pointing errors and turbulence, three operating points are considered.
The severity of pointing errors is varied from the high pointing error regime ($\sigma_s=0.35$~m) to the low pointing error regime ($\sigma_s=0.2$~m). 
The severity of turbulence is constrained to $\sigma_R^2\leq 1$, which corresponds to the weak turbulence regime~\cite[Fig.~7.4]{And05}. 
\begin{figure}[t]
    \begin{minipage}{0.56\textwidth}  
        \centering
        \vspace{-1ex}
            \begin{tikzpicture}[scale=1]    

\definecolor{my_pink}{rgb}{1,0.07,0.65}
\definecolor{my_g}{rgb}{0.09,0.71,0.14}
\definecolor{FEC}{rgb}{0.5, 0.0, 0.13}

\definecolor{fuchsia}{rgb}{0.6,0.4,0.8}

\begin{semilogyaxis}[ 
    width=1\textwidth,
    height=3.3in,   
    xmin=-4, xmax=16,
    ymin=1e-5, ymax=3e-1,  
    xlabel={Transmitted Power, $P$~[dBm]}, 
    ylabel shift=-4ex,
    xlabel shift=-4ex,
    font=\small,
    ylabel={Average BER,~$\bar{P_\text{b}}$}, 
    xtick={-4,-2,...,16},    
    grid=major,
    grid style={dashed,lightgray!75},
]

\addplot [color=black,solid,line width=1pt,mark options={solid}]coordinates {
         (10,1) (10,2)};\label{Orig_var1_jit12_black}
\addplot [color=black,dashed,line width=1pt,mark options={solid}]coordinates {
         (10,1) (10,2)};\label{Dashed_black}
\addplot [color=black,dotted,line width=1pt,mark options={solid}]coordinates {(10,1) (10,2)};\label{dotted_black}
 \addplot [only marks, color=black, mark=*, mark options={solid,scale=0.8, fill=white}]coordinates {(0.35,1)};\label{MC}
\addplot [color=FEC, very thick]
coordinates {(10,1) (10,2)}; \label{FEClimit}

  \addplot [color=FEC, very thick]
   coordinates {(-10,3.84e-3) (16,3.84e-3)};
  

\addplot [color=my_g,solid,line width=1pt,mark options={solid}]file{./txtData/BER_3km/BER_orig_varR09_jit2.txt};

\addplot [ color=my_g,dashed,line width=1pt,mark options={solid}] file {./txtData/ICTON/Aprox2_varR09_jit2.txt};

\addplot [color=my_g,dotted,line width=1pt,mark options={solid}] file{./txtData/UpperBound/Aprox2_varR09_jit2.txt};

\addplot [only marks, color=my_g, mark=*, mark options={solid,scale=0.8, fill=white}]file{./txtData/MC_3km/BER_MC_varR09_jit2.txt};\


\addplot [color=my_pink,solid,line width=1pt,mark options={solid}]file{./txtData/BER_3km/BER_orig_varR01_jit35.txt};\label{Orig_var01_jit35}

\addplot [ color=my_pink,dotted,line width=1pt,mark options={solid}] file{./txtData/UpperBound/varR01_jit35/OOK_aprox2.txt};
    
\addplot [color=my_pink,dashed,line width=1pt,mark options={solid}] file{./txtData/ICTON/Aprox2_varR01_jit35_new.txt};


\addplot [only marks, color=my_pink, mark=*, mark options={solid,fill=white,scale=0.8}]file{./txtData/MC_3km/BER_MC_orig_varR01_jit35.txt};\label{MC_orig_var01_jit35}


\addplot [color=orange,solid,line width=1pt,mark options={solid}] file {./txtData/BER_3km/BER_orig_varR05_jit25.txt};\label{Orig_var05_jit25}

\addplot [color=orange,dotted,line width=1pt,mark options={solid}] file{./txtData/UpperBound/varR05_jit25/OOK_aprox2.txt};
    
\addplot [color=orange,dashed,line width=1pt,mark options={solid}] file{./txtData/ICTON/Aprox2_varR05_jit25.txt};

\addplot [only marks, color=orange, mark=*, mark options={solid,scale=0.8,fill=white}]file{./txtData/MC_3km/BER_MC_orig_varR05_jit25.txt};\label{MC_orig_var05_jit25}

\end{semilogyaxis} 

\node [draw,fill=white,anchor= south west,font=\scriptsize] at (5.35,5.1) {\shortstack[l]{ 
\ref{Orig_var1_jit12_black} \hspace{0.01cm} \eqref{eq:BER_LN}: Exact \\
 \ref{dotted_black} \hspace{0.01cm}  \eqref{eq:BER_Approx2_UpperBound}: \textbf{This work}\\
\ref{Dashed_black} \hspace{0.01cm}  \eqref{eq:BER_Aprox2_ICTON}: \cite{ICTON}\\
\hspace{0.17cm}   \ref{MC} \hspace{0.3cm}  Monte Carlo\\
  \ref{FEClimit} \hspace{0.01cm}  HD-FEC limit}};

\node [coordinate](input) {};

\draw[-,dashed, black,line width=0.5pt] ($(input.north)+(2.03,11.1em)$) -- ($(input.north)+(2.03,13em)$);
\draw[-,dashed, black,line width=0.5pt] ($(input.north)+(2.42,11.1em)$) -- ($(input.north)+(2.42,13em)$);
 \node at ($(input.north) + (2.15, 13.5 em)$) [ scale=1, font=\scriptsize] {$0.94$~dB};

\draw[-,dashed, black,line width=0.5pt] ($(input.north)+(4.5,11.1em)$) -- ($(input.north)+(4.5,13em)$);
\draw[-,dashed, black,line width=0.5pt] ($(input.north)+(4.76,11.1em)$) -- ($(input.north)+(4.76,13em)$);
 \node at ($(input.north) + (4.6, 13.5em)$) [ scale=1, font=\scriptsize] {$0.7$~dB};

\draw[-,dashed, black,line width=0.5pt] ($(input.north)+(3.4,11.1em)$) -- ($(input.north)+(3.4,13em)$);
\draw[-,dashed, black,line width=0.5pt] ($(input.north)+(3.58,11.1em)$) -- ($(input.north)+(3.58,13em)$);
 \node at ($(input.north) + (3.25, 13.5 em)$) [scale=1, font=\scriptsize] {$0.4$ dB};

\draw[-,dashed, black,line width=0.5pt] ($(input.north)+(2.03,11.1em)$) -- ($(input.north)+(2.03,9.3em)$);
\draw[-,dashed, black,line width=0.5pt] ($(input.north)+(2.11,11.1em)$) -- ($(input.north)+(2.11,9.3em)$);
\node at ($(input.north) + (1.6, 8.9 em)$) [ scale=1, font=\scriptsize] {$\Delta\approx0.21$~dB};

\draw[-,dashed, black,line width=0.5pt] ($(input.north)+(4.5,11.1em)$) -- ($(input.north)+(4.5,9.3em)$);
\node at ($(input.north) + (4.8, 8.9em)$) [scale=1, font=\scriptsize] {$\Delta\approx0.08$~dB};

\draw[-,dashed, black,line width=0.5pt] ($(input.north)+(3.4,9.3em)$) -- ($(input.north)+(3.4,11.2em)$);
\node at ($(input.north) + (3.3, 8.9 em)$) [scale=1, font=\scriptsize] {$\Delta\approx0.1$ dB};

\draw[stealth-,solid, black,line width=1pt,rounded corners]($(input.north)+(3.5,16em)$) -- ($(input.north)+(1.1,14.5em)$);
\node at ($(input.north) + (4.2, 16.7 em)$) [scale=1.2, font=\scriptsize] {Increasing};
\node at ($(input.north) + (4.2, 16 em)$) [scale=1.2, font=\scriptsize] {Turbulence};

\node at ($(input.north) + (1.75, 5.9em)$) [scale=1.2, font=\scriptsize] {High pointing};
\node at ($(input.north) + (1.9, 5.1em)$) [scale=1.2, font=\scriptsize] {errors};
\node at ($(input.north) + (7.15, 5.9em)$) [scale=1.2, font=\scriptsize] {Low pointing};
\node at ($(input.north) + (7.2, 5.1em)$) [scale=1.2, font=\scriptsize] {errors};
\draw[stealth-,solid, black,line width=1pt,rounded corners] ($(input.north)+(6.7,5.4em)$) -- ($(input.north)+(2.6,5.4em)$);

\node[ellipse,line width=0.75pt, draw,minimum width = 0.6cm, 
	minimum height = 0.05cm, rotate=45] (e) at (3.25,1) {};
\node at ($(input.north) + (2.4, 2em)$) [scale=1, font=\scriptsize] {$\sigma_s=0.35$ m};
\node at ($(input.north) + (2.5, 1.3em)$) [scale=1, font=\scriptsize] {$\sigma_R^2=0.1$};
 
\node[ellipse,line width=0.75pt, draw,minimum width = 0.6cm, 
	minimum height = 0.05cm,rotate=45] (e) at (7.1,1) {};
\node at ($(input.north) + (6.3, 2em)$) [scale=1, font=\scriptsize] {$\sigma_s=0.2$ m};
\node at ($(input.north) + (6.3, 1.3em)$) [scale=1, font=\scriptsize] {$\sigma_R^2=0.9$};

\node[ellipse,line width=0.75pt, draw,minimum width = 0.6cm, 
	minimum height = 0.05cm,rotate=45] (e) at (5.4,1) {};
\node at ($(input.north) + (4.6,1.7em)$) [scale=1, font=\scriptsize] {$\sigma_s=0.25$ m};
\node at ($(input.north) + (4.6, 1em)$) [scale=1, font=\scriptsize] {$\sigma_R^2=0.5$};
 
\end{tikzpicture}
        \vspace{-4.5ex}
        \caption{Average BER versus transmitted power considering clear air, $L=3$~km, and three different cases for ($\sigma_s$, $\sigma_R^2$): ($0.2,0.9$), ($0.25,0.5$), and ($0.35,0.1$).}
        \label{Fig:Main}
    \end{minipage}
 \hspace{1ex}
    \begin{minipage}{0.4\textwidth}  
        \small
         \vspace{-15ex}
        \captionof{table}{Simulation Parameters}
        \vspace{-7pt}
        \renewcommand{\arraystretch}{1.2}
        \setlength{\tabcolsep}{2.5pt}  
        \begin{tabular}{cc}
        \hline
        \multicolumn{2}{c}{\cellcolor{black!15}{\textbf{FSO Parameters}}} \\
            \hline
           Parameter & Value\\
            \hline
            Wavelength ($\lambda$)    &   $1550$ nm\\
            Link Length ($L$)    &   $3$ km\\
            Noise Standard Deviation ($\sigma_n$)& $10^{-7}$ A/Hz\\  
            Responsivity ($\eta$) & $0.5$ A/W\\
            Beam Waist ($\omega_z$)&   $1.98$ m\\
            Receiver Radius ($a$)  &   $5$~cm\\
            \hline
    \multicolumn{2}{c}{\cellcolor{black!15}{\textbf{Turbulence, Pointing Error and Climatic Parameters}}} \\
             \hline
           Parameter & Value\\
            \hline
             Attenuation Coefficient ($\sigma$)  & $0.2208$~dB/km\\
              Rytov Variance ($\sigma_R^2$) &   $\leq 1$\\ 
            Jitter Angle ($\theta_{s}$) & $0.067-0.116$ mrad \\
            \hline
        \end{tabular}
        \label{tab:sim_param}
    \end{minipage}
    \vspace{-4ex}
\end{figure}
The first ($\sigma_s=0.35$~m and $\sigma_R^2=0.1$), the second ($\sigma_s=0.25$~m and $\sigma_R^2=0.5$), and third ($\sigma_s=0.2$~m and $\sigma_R^2=0.9$) cases are shown in Fig.~\ref{Fig:Main} with pink, orange, and green, resp.
Solid, dotted, and dashed curves correspond to the exact average BER~\eqref{eq:BER_LN}, the approximate average BER in~\eqref{eq:BER_Approx2_UpperBound}, and the approximate average BER in~\eqref{eq:BER_Aprox2_ICTON} (from \cite{ICTON}), resp.
BER estimates obtained via Monte Carlo simulations are also shown with circles. 
Moreover, the so-called HD forward error correction (FEC) threshold of $3.84\times10^{-3}$ for a coding rate 0.937~\cite[Table I]{FEC} is also shown.
Finally, we define an accuracy parameter $\Delta$ as the difference in transmit power $P$ at which the exact BER~\eqref{eq:BER_LN} and our proposed approximation~\eqref{eq:BER_Approx2_UpperBound} reach the HD FEC limit.

Note from Fig.~\ref{Fig:Main} that the best BER performance is achieved in the first case (pink curves), since, although high pointing errors are considered, the dominant effect in this study is atmospheric turbulence.

We observe from Fig.~\ref{Fig:Main} that for the first case (high pointing errors and very weak turbulence, pink curves), the BER approximation~$\eqref{eq:BER_Aprox2_ICTON}$ proposed in~\cite{ICTON} is $0.94$~dB away from the exact BER~\eqref{eq:BER_LN} at the HD FEC limit.
For the second (medium pointing errors and weak turbulence, orange curves) and third cases (the lowest pointing errors and highest weak turbulence, green curves), our previous approximation~\eqref{eq:BER_Aprox2_ICTON} still results in imprecise estimates, off by $0.4$~dB and $0.7$~dB at the HD-FEC threshold, respectively.
The new approximation~\eqref{eq:BER_Approx2_UpperBound} we proposed in this paper, however, has a gap of only $\Delta=0.21$~dB to the actual BER, for the first case, while for the second and third cases, the gap $\Delta$ is below $0.1$~dB. 
Additionally, we see in Fig.~\ref{Fig:Main} that for the BER values of interest (i.e., BERs between $10^{-2}-10^{-5}$), operating in the high pointing error regime, the slope of the BER approximation in~\eqref{eq:BER_Aprox2_ICTON} (pink dashed curve) differs from the actual slope. 
Our new approximation~\eqref{eq:BER_Approx2_UpperBound}, however, provides BER curves with correct slopes.

\vspace{-1ex}
\section{Conclusions}
\vspace{-1ex}
We developed a simple analytical expression for the average BER for OOK signaling over an FSO channel considering weak atmospheric turbulence, pointing errors, and atmospheric losses.
Our proposed expression provides a better trade-off in terms of complexity vs. accuracy to those available in the literature. The proposed BER approximation is shown to be in close agreement with the true BER. Moreover, our approximation obtains the correct slope of the transmit power vs. BER curve which was a shortcoming of a previously-proposed approximation.
Future work includes the experimental validation of the results presented and the extension of this study to symbol error rates and generalization to higher-order modulation formats. 

\vspace{1ex}
{\centering \normalsize\bfseries\scshape Acknowledgement\\} 
\vspace{0.5ex}
This publication is part of the project BIT-FREE with file number 20348 of the research programme Open Technology Programme which is (partly) financed by the Dutch Research Council (NWO).

\end{document}